\documentclass[useAMS,usenatbib]{mn2e}

\newcommand{\be}{\begin{eqnarray}}
\newcommand{\ee}{\end{eqnarray}}

\usepackage{graphicx}
\usepackage{amssymb}
\usepackage{amsmath}
\title[Production of gamma rays and neutrinos in SS433]{Production of gamma rays and neutrinos in the dark jets
of the microquasar SS433}
\author[M. M. Reynoso, G. E. Romero, and H. R. Christiansen]{M. M. Reynoso$^{1}$\thanks{E-mail:
mreynoso@mdp.edu.ar (MMR)} \thanks{Fellow of CONICET}, G. E.
Romero$^{2}$
\thanks{Member of CONICET}, and H. R. Christiansen$^{3}$\\
$^{1}$Departamento de F\'{\i}sica, Facultad de Ciencias Exactas y
Naturales, Universidad Nacional
de Mar del Plata,\\ Funes 3350, Mar del Plata, 7600, Argentina\\
$^{2}$Instituto Argentino de Radioastronom\'{\i}a, CONICET, C.C.5,
Villa Elisa, 1894, Argentina and \\ Facultad de Ciencias
Astron\'omicas y Geof\'{\i}sicas, Universidad Nacional
 de La Plata, Paseo del Bosque, La Plata, 1900, Argentina \\
$^{3}$State Univesity of Cear\'a, Physics Dept., Av. Paranjana 1700,
60740-000 Fortaleza - CE, Brazil}
\begin{document}

\date{Accepted 2008 April 20. Received 2008 April 02; in original form 2008 January 18}


\maketitle

\label{firstpage}

\begin{abstract}
We study the spectral energy distribution of gamma rays and
neutrinos in the precessing microquasar SS433 as a result of $pp$
interactions within its dark jets.
Gamma-ray absorption due to interactions with matter of the extended
disk and of the star is found to be important, as well as absorption
caused by the UV and mid-IR radiation from the equatorial
envelopment. We analyze the range of precessional phases for which
this attenuation is at a minimum and the chances for detection of a
gamma-ray signal are enhanced.
The power of relativistic protons in the jets, a free parameter of
the model, is constrained by HEGRA data. This imposes limits on the
gamma-ray fluxes to be detected with instruments such as GLAST,
VERITAS and MAGIC II. A future detection of high energy neutrinos
with cubic kilometer telescopes such as IceCube would also yield
important information about acceleration mechanisms that may take
place in the dark jets. Overall, the determination of the ratio of
gamma-ray to neutrino flux will result in a key observational tool
to clarify the physics of heavy jets.

\end{abstract}


\section{Introduction}
The famous and enigmatic microquasar SS433 has been matter of
investigation for more than two decades. Consisting of a donor star
feeding mass to a black hole, it presents two oppositely directed,
precessing jets with hadronic content\footnote{Iron lines with a
shift corresponding to a velocity of $v\sim 0.26 c$ have been
detected, for instance, by Migliari et al. (2002).}. We refer to the
relativistic collimated outflows as `dark' jets \citep{gallo} since
the very high kinetic luminosity $L_{\rm k}\sim 10^{39}$ erg
s$^{-1}$ \citep{dubner} appears as the dominant power output of the
ejected material, having imprinted a deformation on the supernova
remnant W50. 

Most of the radiative output of the system is observed in the UV and
optical bands, whereas the X-ray emission detected is $\sim 1000$
lower than the kinetic energy of the jets, probably due to a
screening effect with an equatorial outflow
\citep{fabrikaXrays,marshall}. The gamma-ray emission above $0.8$
TeV has been constrained by HEGRA to be $\Phi_\gamma<8.93 \times
10^{-13} {\rm cm}^{-2}{\rm s}^{-1}$ \citep{HEGRA} whereas the
neutrino flux upper limit according to AMANDA-II data is
$\Phi_\nu<0.21 \times 10^{-8} {\rm cm}^{-2}{\rm s}^{-1}$
\citep{Halzen06}.

In previous hadronic models for high energy emission from
microquasars, relativistic protons in the jets interact with target
protons from the stellar wind of the companion star
\citep{rom03,hugo02,Or07}. Since in the case of SS433 there is no
evidence of such a strong stellar wind, in this work we investigate
the possible production of gamma rays and neutrinos resulting from
$pp$ interactions between relativistic and cold protons within the
jets themselves.

\section[]{Preliminaries}

\begin{figure}
\includegraphics[trim = 0mm 0mm 0mm 0mm, clip, width=8cm,angle=0]{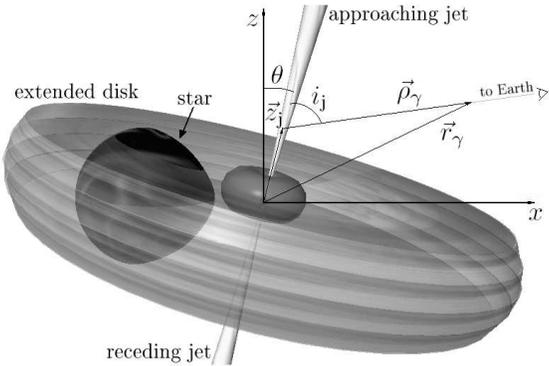}
\caption{Schematic view of the SS433. The \textit{approaching} jet
is most of the time closest to our line of sight and the
\textit{receding} one is oppositely directed.} \label{Figrender}
\end{figure}

The binary SS433, distant $5.5$ kpc from the Earth, displays two
mildly relativistic jets ($v_{\rm b}\approx 0.26$c) that are
oppositely directed and precess in cones of half opening angles of
$\theta\approx 21^\circ$. The line of sight makes an angle
$i=78^\circ$ with the normal to the orbital plane and a
time-dependent angle $i_{\rm j}(t)$ with the approaching jet (see
Fig. {\ref{Figrender}}). Assuming that $\psi(t)$ is the precessional
phase of the approaching jet, we shall follow the convention that
when $\psi=0$ the mentioned jet points closer to the Earth. Then,
when $\psi=0.5$, it has performed half of the precession cycle and
it makes its largest angle with the line of sight. The mass loss
rate in the jets is $\dot{m}_{\rm j}= 5\times 10^{-7}{ M}_\odot {\rm
yr}^{-1}$, the period of precession is $162$ d and the orbital
period is $13.1$ d \citep{Fabrika100}. The donor star and the
compact object are thought to be embedded in a thick expanding disk
which is fed by a wind from the supercritical accretion disk around
the black hole \citep{Zwitter}. This equatorial envelope is
perpendicular to the jets and according to \citet{Fabrika100} we
assume that it has a half opening angle $\alpha_{\rm w} \approx
30^\circ $, a mass loss rate $\dot{M}_{\rm w}\approx 10^{-4}{
M}_\odot{\rm yr}^{-1} $ and a terminal velocity $v_{\rm w}\sim 1500
{\rm \ km \ s}^{-1}$. Also, this extended disk has been recognized
as the origin of both the UV and mid-IR emission
\citep{Gies02a,Fuchs05} which can cause significant absorption of
gamma-rays as discussed in \citet{last}.


The spectral identification of the companion star has been difficult
due to the presence of the extended disk, since the star is often
partially or totally obscured by it. After convenient observations
at specific configurations of precessional and orbital phases it has
became quite clear that the star is an A-supergiant
\citep{Hillwig,Barnes,Chere}. We assume the masses of the components
as derived from INTEGRAL observations \citep{Chere}, $M_{\rm bh}= 9
{ M}_\odot$ and $M_\star= 30 { M}_\odot$ for the black hole and star
respectively. This corresponds to an orbital separation $a \simeq 79
\ { R}_\odot$ for a zero-eccentricity orbit as it is the case for
SS433. Since the star is believed to fill its Roche lobe, the
implied radius according to \citet{Eggleton} is $R_L\simeq 38 {
R}_\odot$.



\subsection{Outline of the jet model}
\textbf{} We assume that a magneto-hydrodynamic mechanism for jet
ejection operates in SS433, that is, ejection is realized through
the conversion of magnetic energy into matter kinetic energy. The
magnetic energy density is supposed to be in equipartition with the
kinetic energy density of the ejected particles, so that the
corresponding magnetic field along the jet is given by
 \be
 B(z_{\rm j})= \sqrt{8\pi e_{\rm j}},
 \ee
where the kinetic energy density is
 \be
 e_{\rm j}=\frac{\dot{m}_{\rm j}E_{\rm k}}{m_{\rm p}v_{\rm b}\pi R_{\rm
 j}^2(z_{\rm j})}.
 \ee
Here $E_{\rm k}$ is the classical kinetic energy of a jet proton
with velocity $v_{\rm b}$ and $R_{\rm
 j}(z)$ is the jet radius at the height $z_{\rm j}$ along the jet axis.

The jets are modeled as cones with a half opening angle $\xi_{\rm
j}\approx 0.6^\circ$ \citep{marshall}. Assuming an initial jet
radius $R_0=R_{\rm j}(z_0)\approx 5 R_{\rm Sch}$, where $R_{\rm
Sch}=2GM_{\rm bh}/{c^2}$, we find the injection point as $z_0=
R_0/\tan{\xi_{\rm j}}\simeq 1.3 \times 10^{9}$cm. Since the jets are
heavy as compared to other similar objects, it is reasonable to
admit that they are cold matter dominated. In this case, we assume
that a small fraction of relativistic or hot particles are confined
by the cold plasma. According to \citet{broadband} the pressure of
cold particles is greater than that of the relativistic ones if the
ratio of cold to hot particles is less than 1/1000, and this
condition will be greatly satisfied provided that the luminosity
carried by relativistic particles is required to be smaller than the
total kinetic luminosity of the jet.

Particle acceleration is supposed to take place via diffusive
acceleration by internal shocks converting bulk kinetic energy into
random kinetic energy. According to the standard model for
non-relativistic shock acceleration (e.g. Blandford \& Eichler 1987
and references therein) we expect that the relativistic proton
spectrum is given by a power-law, $N'_p({E'}_p)= K_0
{E'}_p^{-\alpha}$ at $z_{\rm j}=z_0$, where the spectral index is
the standard value for first order diffusive shock acceleration,
$\alpha=2$. The flux of these protons hence evolves with $z_{\rm j}$
as $$J'_p({E'}_p)= \frac{cK_0}{4\pi} (z_0/z_{\rm j})^2
{E'}_p^{-\alpha}$$ in the jet frame, which transformed to the
observer frame \citep{Purmohammad} reads

\begin{multline}
   J_p(t,E_p,z_{\rm j})=\frac{c K_0}{4 \pi}
     \left(\frac{z_0}{z_{\rm j}}\right)^ 2 \times  \label{Jplab} \\
     \frac{\Gamma^{-\alpha+1} \left(E_p-\beta_{\rm b}
\sqrt{E_p^2-m_p^2c^4} \cos i_{\rm j} \right)^{-\alpha}}{\sqrt{\sin
^2 i_{\rm j} + \Gamma^2 \left( \cos i_{\rm j}(t) - \frac{\beta_{\rm
b} E_p}{\sqrt{E_p^2-m_p^2 c^4}}\right)^2}}  \\ \equiv
\left(\frac{z_0}{z_{\rm j}}\right)^2 \tilde{J}_p(E_p,t),
\end{multline}
where $i_{\rm j}(t)$ is the angle between the jet axis and the line
of sight, $\beta_{\rm b}= 0.26$, and $\Gamma=\left[ 1-\beta_{\rm
b}^2 \right]^{-1/2}$ is the jet Lorentz factor. The normalization
constant $K_0$ is obtained by specifying the fraction of power
carried by the relativistic protons, $q_{\rm rel}$,
 \be
 \pi R_0^2\int_{E_p'^{\rm(min)}}^{E_p'^{\rm(max)}} J'_p(E'_p) E'_p dE'_p= q_{\rm
 rel}L_k,
 \ee
so that
 \be
 K_0 = \frac{4 q_{\rm rel} L_k}{c  R_0^2\ln
 \left(\frac{{E'}_{p}^{\rm(max)}}{{E'}_p^{\rm(min)}}\right)},
 \label{K0norm}
 \ee
where we take ${E'}_p^{\rm(min)}\approx 1$ GeV and the maximum
proton energy ${E'}_p^{\rm(max)}$ will be determined in the next
section. We shall adopt, for the illustrative predictions of
neutrino and gamma-ray fluxes, a tentative value $q_{\rm rel}=
10^{-4}$, but a full discussion of the possible range for this
parameter will be presented in Sect. \ref{neutrinos}.

\section{Hadronic processes in the jets}

Relativistic protons in the jets are subject to different mechanisms
that can make them lose energy. In this section we analyze the
energy range where $pp$ collisions are the dominant cooling process
that will produce the corresponding gamma rays and neutrinos in
SS433.

 \subsection{Acceleration}

The acceleration rate of protons up to an energy $E_p$ can be
estimated as \citep{BRS90}:
 \be
t_{\rm accel}^{-1} \approx \eta \frac{ceB}{E_p},
 \ee
where $\eta\sim \beta_{\rm b}^2$ is the acceleration efficiency.
 Clearly, as long as the latter rate is greater than the total loss
rate for a given energy, protons will be effectively accelerated up
to that energy.

As mentioned in Sect. 2.1, we assume that the protons are
accelerated at shocks produced by collisions of plasma outflows with
different bulk velocities. In the frame of the shock, the
conservation equations imply that the upstream velocity is
significantly higher than the downstream velocity, i.e., $v_{\rm
u}/v_{\rm d}= (\gamma_{\rm heat}+1)/(\gamma_{\rm heat}-1)= \xi$,
with $\gamma_{\rm heat}$ the ratio of specific heats and $\xi$ the
compression factor. In this way, the two regions may be regarded as
two converging flows. The Fermi first order acceleration mechanism
then operates to produce a power law particle spectrum, which is
essentially independent of the microphysics involved. Strong shocks
($\xi\sim 4$) can be non-relativistic as it is the case, for
instance, in supernova remnants and colliding wind massive binaries.
In the case of sub-relativistic jets, strong shocks are expected as
suggested by the non-thermal synchrotron radio spectra observed from
the jets of microquasars (Fender 2004). Non-linear effects like
shock modification by the pressure of the relativistic particles or
magnetic field effects can result in a variety of spectral indexes.
The reader is referred to the recent paper by Rieger et al. (2006)
on Fermi acceleration in astrophysical jets, which includes a
section on mildly relativistic microquasar outflows.

The acceleration of protons proceeds only for protons with a
threshold energy that allows the diffusive acceleration process to
take place (Rieger et al. 2006). Then, only the supra-thermal tail
of the Maxwellian distribution of cold protons will be affected by
the process (Bosch-Ramon et al. 2006). This has the result that just
a small fraction of the total power carried by the jet is converted
to relativistic particles.

 \subsection{Cooling rates and maximum particle energy}

  The density of cold protons at a distance $z_{\rm j}$ from the black hole in each
jet is
 \be
n_p(z_j)\simeq \frac{\dot{m}_{\rm j}}{\pi [R_j(z_j)]^2 m_p v_{\rm
b}} \label{npcold}.
 \ee
 These cold protons serve as targets for the relativistic ones, so that
 the rate of
$pp$ collisions in the jet is given by
 \be
t_{pp}^{-1}= n_p(z_{\rm j}) c \sigma_{pp}^{\rm(inel)}(E_p)K_{pp},
 \ee
where the inelasticity coefficient is taken to be $K_{pp}\approx
1/2$ since on average, the leading proton losses half of its
total energy per collision. 

The cross section for inelastic $pp$ interactions can be
approximated by \citep{Kelner06}
 \be
\sigma_{pp}^{\rm(inel)}(E_p)= (34.3+ 1.88 L+ 0.25 L^2)\times \\
 \left[1-\left(\frac{E_{\rm th}}{E_p}\right)^4 \right]^2 \
\times 10^{-27}{\rm cm}^2,
 \ee
where $L= \ln(E_p/1000{\rm \ GeV})$ and $E_{\rm th}=1.22 {\rm \
GeV}$.

Cooling by $p\gamma$ interactions can take place mainly via
photomeson production ($\gamma p \rightarrow p \pi^i $) and pair
production ($ \gamma p\rightarrow p e e^+$) \citep{BRS90}. The
corresponding cooling rate can be obtained from  \citep{AD03}

 \be
t_{p\gamma}^{-1}=\int_{\frac{E'_{\rm th}}{2\gamma_p}}^\infty dE
\frac{c  n_{\rm ph}(E)}{2\gamma_p^2E^2} \int_{E_{\rm th}}^{2\gamma_p
E} dE_r \sigma_{p\gamma} K_{p\gamma} E_r dE_r, \label{tpg}
 \ee
where $E'_{\rm th}\approx 150 \ {\rm MeV}$, $\gamma_p$ is the
Lorentz factor of the proton, $n_{\rm ph}(E)$ represents the density
of target photons, $\sigma_{p\gamma}$ will be the inelastic cross
section appropriate for photopion and photopair creation, and
$K_{p\gamma}$ is the corresponding inelasticity coefficient.

Photopion production will occur when protons collide with X-ray
photons, for which, based on  \citet{Chere}, we adopt a
Bremsstrahlung X-ray distribution for $2 {\rm \ keV}< E < 100$ keV,
 \be
  n_{\rm X}(E)=L_{\rm X} \frac{e^{-{E}/{(kT_e)}}}{4\pi z_{\rm j}^2 E^{2}}
\  ({\rm erg}^{-1}{\rm cm}^{-3}),
  \label{Xrays}
  \ee
where $kT_e \approx 30$ keV and $L_X= 10^{36} \ {\rm erg} \ {\rm
s}^{-1}$. These X-ray photons are considered to be originated in a
corona surrounding the inner accretion disk, as suggested in
\citet{Chere}.

The cross section for photopion production is approximated by
\citep{AD03}
 \begin{multline}
 \sigma_{p\gamma}^{(\pi)}=  \Theta(E_r-200 \ {\rm
MeV})\Theta(500 \ {\rm MeV}-E_r) \ 3.4\times 10^{-28}{\rm cm}^2 \\
+ \Theta(E_r-500 \ {\rm MeV})\ 1.2\times 10^{-28}{\rm cm}^2,
\label{sigphotopion}
 \end{multline}
and the inelasticity coefficient for photopion production is
 \begin{multline}
 K_{p\gamma}^{(\pi)}=  \Theta(E_r-200{\rm \
MeV})\Theta(500{\rm \ MeV}-E_r) \ 0.2 \\
+ \Theta(E_r-500{\rm \ MeV})\ 0.6.
 \end{multline}

The contribution of the $e^-e^+$ pair creation process to the total
$p\gamma$ cooling rate is calculated also using Eq. (\ref{tpg}), but
the soft photon density in this case includes also the contribution
associated with the UV emission from the extended disk,

$$n_{\rm
ph}(E)=n_{\rm UV}(E,\Omega){\frac{\pi R_{\rm out}^2}{z_{\rm j}^2}}+
n_{\rm X}(E).$$

Based on the discussion in \citet{Gies02a}, the UV photons with
wavelengths in the range ($1000{\rm \ \AA},10000{\rm \ \AA}$)
correspond to a blackbody distribution  with $T_{\rm UV}=21000$ K.
Hence, we take the corresponding radiation density as
 \be
 n_{\rm UV}(E,\Omega)= {2E^2}{(h c)^{-3}( e^{E/kT_{\rm UV}}-
1)^{-1}}.
 \ee

For this process we consider the expressions for cross section and
inelasticity given in \citet{BRS90}:
 \begin{multline}
\sigma_{p\gamma}^{(e)}= 5.8\times 10^{-28}{\rm cm}^2\left[3.11L'- 8.07 + \right.\\
\left. \left(2m_e c^2/{E}\right)^2\left(2.7 L'- L'^2 + 0.67L'^3+0.55
\right)-\right. \label{sigphotopair} \\ \left. \left({2m_e
c^2}/{E}\right)^4\left(0.19L'+0.13 \right)-\left({2m_e
c^2}/{E}\right)^6\left(0.01L'\right) \right] 
 \end{multline}
 and
 \be
K_{p\gamma}^{(e)}= 4\frac{m_e^2c^2}{m_p E_r}\left[\frac{-8.78+ 5.51
\
 L'- 1.61 \ L'^2+ 0.69 \ L'^3}{3.11\ L'- 8.07}\right]
 \ee
with $L'=\ln\left(\frac{2E}{m_e c^2}\right)$.


The accelerated protons can also lose energy in the form of
synchrotron radiation at a rate
 \be
t_{\rm sync}^{-1}=\frac{4}{3}\left(\frac{m_e}{m_p}\right)^3\frac{
\sigma_{\rm T}B^2}{m_e c \ 8\pi}\gamma_p \label{tsyn},
 \ee
and via Inverse Compton scatterings with X-ray and UV photons at a
rate
 \be
t_{\rm IC}^{-1}=\frac{4}{3}\left(\frac{m_e}{m_p}\right)^3\frac{
\sigma_{\rm T}e_{\rm ph}}{m_e c}\gamma_p,
 \ee
where $$e_{\rm ph}=\int_{E_{\rm min}}^{m_p^2c^4/E_p} n_{\rm ph}(E)\
E\ dE$$ is the corresponding density of energy in target soft
photons \citep{BRS90}.

It is also expected that the accelerated protons will suffer
adiabatic losses because of the expansion undergone by the jets. The
corresponding adiabatic cooling rate can be written as (see e.g.
Bosch Ramon et al. 2006)
 \be
t_{\rm adiab}^{-1}= \frac{2}{3}\frac{v_{\rm b}}{z_{\rm j}}
 \ee

\begin{figure}
\includegraphics[trim = 7mm 5mm 0mm 0mm, clip, width=8cm,angle=0]{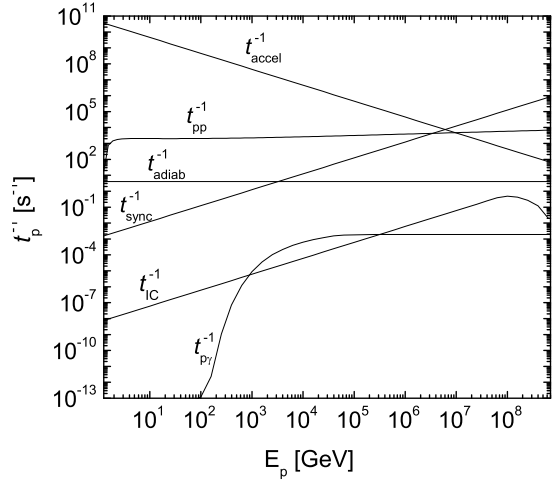}
\caption{Proton accelerating and cooling rates at the base of the
jets.} \label{Figtploss}
\end{figure}

We show the obtained results for the acceleration and cooling rates
at the base of each jet ($z_{\rm j}=z_0$) in Fig. \ref{Figtploss}.

As it can be seen from this plot, the expected cooling rates for
$p\gamma$ and Inverse Compton interactions are found to be
negligible. We infer that the protons which can be effectively
accelerated up to energies below
 \be
  E_p^{(\rm max)} \approx 3.4 \times 10^6 {\rm GeV}, \label{Epmax}
 \ee
will cool efficiently mainly via $pp$ collisions at $z_0$. For
higher energies, synchrotron losses become dominant. On the other
hand, as $z_{\rm j}$ increases, adiabatic losses also grow and the
maximum proton energy at which $pp$ collisions dominate can change
along the jets.

\begin{figure}
\includegraphics[trim = 7mm 5mm 0mm 0mm, clip, width=8cm,angle=0]{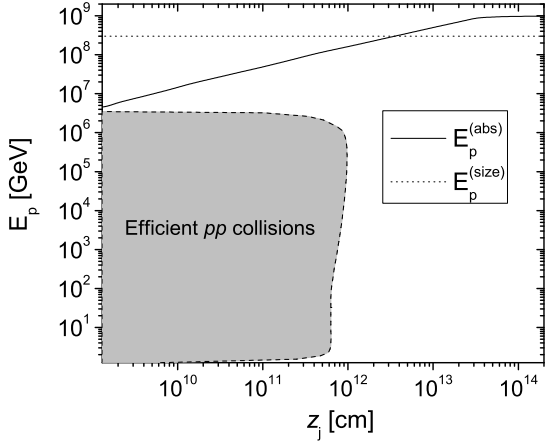}
\caption{The region in the $(z_{\rm j},E_p)$ plane where $pp$
interactions are the dominant cooling mechanism is indicated in
gray. The absolute maximum proton energy ({\it solid line}) and the
maximum proton energy from the size constraint ({\it dotted line})
are also shown.} \label{FigEpmax}
\end{figure}

The absolute maximal energy for protons at a given $z_{\rm j}$,
$E_{p}^{\rm(abs)}$ (see Fig. {\ref{FigEpmax}}) is obtained from
$t^{-1}_{\rm accel}=t^{-1}_{pp}+ t^{-1}_{\rm adiab} + t^{-1}_{\rm
sync} + t^{-1}_{\rm IC}+ t^{-1}_{p\gamma}$.

The size constraint, implying that the proton gyro-radio has to be
smaller than the radius of the jet, i.e.
 $E_p<E_p^{\rm (size)}=e R_{\rm j} B \approx 3\times 10^8 {\rm
 GeV}$,
does not happen to limit the energy of the protons at the bases of
the jets. Note also that we have the same value of $E_p^{\rm(size)}$
for larger values of $z_{\rm j}$ along the jets because $B\propto
R_{\rm j}^{-1}$. Nevertheless, the size constraint will limit the
energy of the accelerated protons at distances $z_{\rm j}\gtrsim
3\times 10^{12}$cm from the black hole (see Fig. \ref{FigEpmax}).

In the $(z_{\rm j},E_p)$ region where $pp$ collisions dominate the
cooling mechanism, the condition $t_{pp}^{-1}> t_{\rm sync}^{-1}+
t_{\rm adiab}^{-1}+t_{\rm IC}^{-1} + t_{p\gamma}^{-1}$ must hold.
This region is indicated in the shaded zone of Fig. \ref{FigEpmax}.
It can be seen that the maximum energies for efficient cooling
through $pp$ interactions are $E_p^{\rm(max)}\sim 3\times 10^6$ GeV
for $z_{\rm j}<z_1$, where $z_1\approx 10^{12}$cm.\footnote{Since
the necessary maximal energy in equation (\ref{K0norm}) has to be
expressed in the frame comoving with the jet, we take
${E'}_p^{\rm(max)}\simeq \Gamma(E_p^{\rm(max)}-\beta_{\rm
j}\sqrt{{E_p^{\rm(max)}}^{2}- m_p^2c^4}$), where
$E_p^{\rm(max)}\approx 3.4\times 10^{6}$GeV at the base of the jet.}
Therefore, the jet will become essentially cold and observationally
dark, unless some re-acceleration mechanism could operate (e.g.
mediated by re-collimation shocks or terminal shocks).

It is also worth noting that decreasing the parameter $q_{\rm rel}$
does not change the maximum proton energy given by (\ref{Epmax})
since in that case the target proton density is essentially
unchanged and so is the $pp$ cooling rate. This allows us to take
$q_{\rm rel}$ as a free parameter that can be easily factored out in
all of our predictions for gamma-ray and neutrino signals. \\


 \subsection{High-energy gamma rays from $pp$ interactions}

The collision of a certain number of high-energy protons with cold
protons in the jets will cause the production of secondary gamma
rays and neutrinos. Following the treatment of \citet{Kelner06},
which is based on SIBYLL simulations of $pp$ interactions including
perturbative QCD effects, the spectrum of produced gamma-rays with
energy $E_\gamma= x E_p$ for a primary proton with energy $E_p$
reads
 \be
F_\gamma(x,E_p)= B_\gamma\frac{\ln
x}{x}\left(\frac{1-x^{\beta_\gamma}}{1+ k_\gamma x^{\beta_\gamma}(1-
x^{\beta_\gamma})} \right)^4 \times  \nonumber \\
\left[\frac{1}{\ln x}-\frac{4\beta_\gamma
x^{\beta_\gamma}}{1-x^{\beta_\gamma}}- \frac{4 k_\gamma \beta_\gamma
x^{\beta_\gamma}(1-2x^{\beta_\gamma})  } { 1+k_\gamma
x^{\beta_\gamma} (1-x^{\beta_\gamma})} \right],
\label{KelnerGammaHigh}
 \ee
where
 \be
B_\gamma &=& 1.3+ 0.14  \ L+ 0.011  \ L^2\\
\beta_\gamma &=& \frac{1}{1.79+ 0.11 \ L+ 0.008 \ L^2}\\
k_\gamma &=& \frac{1}{0.801+ 0.049 \ L+ 0.014 \ L^2},
 \ee
with $L=\ln{(E_p/1{\rm \ TeV})}$.




For $E_\gamma>100$ GeV, we shall consider the gamma-ray emissivity
 at a height $z_{\rm j}$ along the jets as
 \be
\frac{dN_\gamma(t,E_\gamma,z_{\rm j})}{dE_\gamma}= \int_{x_{\rm
min}}^{x_{\rm max}} \sigma_{pp}^{\rm
inel}\left(\frac{E_\gamma}{x}\right)
J_p\left(t,\frac{E_\gamma}{x},z_{\rm j}\right)\times \nonumber \\
F_\gamma\left(x,\frac{E_\gamma}{x}\right)  dx \\  \equiv
\left(\frac{z_0}{z_{\rm
j}}\right)^2\frac{d\tilde{N}_\gamma(t,E_\gamma)}{dE_\gamma}
 \ee
in units ${\rm GeV}^{-1}{\rm s}^{-1}$. The integration limits
$x_{\rm min}$ and $x_{\rm max}$ are chosen in order to cover the
proton energy range where $pp$ collisions dominate at each $z_{\rm
j}$, as shown in Fig. \ref{FigEpmax}. On the other hand, for
$E_\gamma<100$ GeV, we shall consider, as suggested in
\citet{Kelner06}, the emissivity obtained using the
$\delta$-functional approximation
 \be
\frac{dN_\gamma(t,E_\gamma,z_{\rm j})}{dE_\gamma}= 2\int_{E_{\rm
min}}^{E_{\rm max}} \frac{q_\pi(t,E_\pi,z_{\rm j})}{\sqrt{E_\pi^2-
m_\pi^2c^4}}dE_\pi,
 \ee
where $E_{\rm min}= E_\gamma+ \frac{m_\pi^2c^4}{4E_\gamma}$, $E_{\rm
max}= K_\pi(E_p^{\rm(max)}-m_p c^2)$, and
 \begin{multline}
q_\pi(t,E_\pi,z_{\rm j})=\frac{\bar{n}}{K_\pi}\sigma_{pp}^{\rm
inel}(m_p c^2+ \frac{E_\pi}{K_\pi}) \times  \\ J_p(t,m_p c^2+
\frac{E_\pi}{K_\pi},z_{\rm j}).
 \end{multline}
Here, $K_\pi\approx 0.17$ is the fraction of the proton kinetic
energy that is transferred to the gamma rays or leptons. The number
of produced pions, $\bar{n}$, is a free parameter of the model that
is fixed by requiring continuity of the gamma-ray emissivity at
$E_\gamma= 100$ GeV.
\begin{figure}
\includegraphics[trim = 10mm 6mm 0mm 8mm, clip, width=9cm,angle=0]{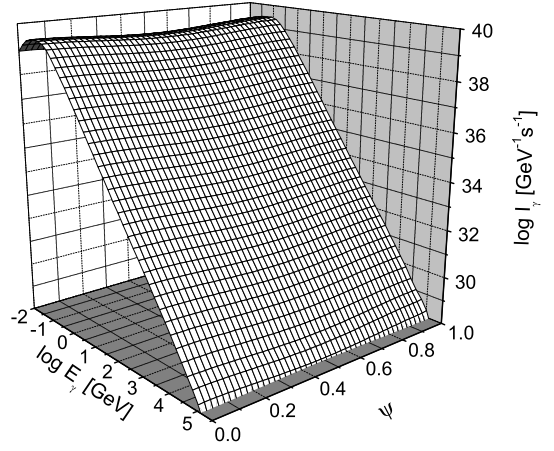}
\caption{Spectral intensity of gamma-rays produced in the
approaching jet as a function of the precessional phase and the
gamma-ray energy.} \label{FigIga-a}
\end{figure}

The spectral intensity of gamma rays emitted from the jet can be
obtained from 
 \be
I_\gamma(t,E_\gamma)&=& \int_{z_0}^{z_1}\pi(z_{\rm j}\tan \xi)^2 n_p
\frac{dN_\gamma(t,E_\gamma,z_{\rm j})}{dE_\gamma} dz_{\rm j}\\
&\simeq& \frac{\dot{m}_{\rm j}z_0}{m_p v_{\rm
b}}\frac{d\tilde{N}_\gamma(t,E_\gamma)}{dE_\gamma}.
 \ee
We show the obtained result for gamma rays produced in the
approaching jet in Fig. \ref{FigIga-a}.


\subsection{High-energy neutrino emission from $pp$ interactions}

Neutrinos are produced by the decay of the charged pions resulting
from $pp$ interactions,
 \begin{eqnarray}
 \pi^-\rightarrow \mu^-\bar{\nu}_\mu \rightarrow e^-\nu_\mu \bar{\nu}_e
 \bar{\nu}_\mu \\
 \pi^+\rightarrow \mu^+ {\nu}_\mu \rightarrow e^+\bar{\nu}_\mu \nu_e {\nu}_\mu
 \end{eqnarray}
In the present context, the equipartition condition implies a strong
magnetic field at the base of the jet. Hence, the charged mesons
will undergo important losses due to synchrotron radiation. In order
to find the local steady distribution of the parent charged pions at
fixed $E_p$, we have considered the following transport equation

 \be
\frac{d b_{\pi}(E_\pi)F_\pi(E_\pi)}{dE_\pi} +
\frac{F_\pi(E_\pi)}{\tau_{\pi}(E_\pi)}= F_\pi^{\rm(inj)}(E_\pi)
t_{pp}^{-1}.  \label{te}
 \ee
Here, $\tau_{\pi}= \tau_{\pi}^{(0)}\gamma_\pi$, with $
\tau_{\pi}^{(0)}=2.6\times 10^{-8}$s, is the pion mean lifetime,
$b_{\pi}(E_\pi)= E_\pi (t^{-1}_{\rm sync}(E_\pi,z_{\rm j})+
t^{-1}_{\rm
 adiab}(z_{\rm j}))$, and
 \begin{multline}
 F_\pi^{\rm (inj)}(E_\pi)=4\alpha B_\pi x_\pi^{\alpha-1}\left(\frac{1-x_\pi^\alpha}{1-r
 x_\pi^\alpha(1-x_\pi^\alpha)}\right)^4\times \\
 \left(\frac{1}{1-x_\pi^\alpha}+ \frac{r(1-2x_\pi^\alpha)}{1+
 rx_\pi^\alpha(1-x_\pi^\alpha)}\right)\left(1-\frac{m_\pi c^2}{x_\pi
 E_p}\right)^{1/2}
  \end{multline}
is the distribution of injected pions per $pp$ collision
\citep{Kelner06}, where $x_\pi= E_\pi/E_p$, $ B_\pi=a+ 0.25$, $a=
3.67+ 0.83 L+ 0.075 L^2$, $r= 2.6/\sqrt{a}$, and $\alpha=
0.98/\sqrt{a}$. The transport equation (\ref{te}) includes the
effects of decays and energy loss of pions in the left member and
the injection of pions in the right side.

The corresponding solution can be written as
\begin{multline}
 F_\pi(E_\pi)=\int_{E_\pi}^{E_p} \frac{F_\pi^{\rm(inj)}(E') t^{-1}_{pp}}{\left|b_\pi(E_\pi)\right|}\times \\
  \exp\left\{\frac{1}{b_z E_\pi}- \frac{1}{b_z E'} + \frac{a_z}{b_z^2}\log\left(\frac{E_\pi}{E'}\right) \right.+ \\
 \left.   \frac{a_z}{b_z^2}\log\left(\frac{b_z+ a_z E'}{b_z+ a_z E_\pi}\right)  \right\} 
 dE',
 \end{multline}
where
  \be
  a_{z}&=&\frac{4}{3}\left(\frac{m_e}{m_\pi}\right)^3\frac{\sigma_{T}B^2(z_{\rm
j})\tau_\pi^{(0)}}{8\pi \ m_e c \left(m_\pi c^2\right)^2},\nonumber \\
  b_{z}&=& \frac{2}{3}\frac{v_{\rm b}}{z_{\rm j}}
  \frac{\tau_\pi^{(0)}}{m_\pi c^2}.\nonumber
  \ee

The spectrum of high energy neutrinos from the direct decay of the
steady distribution of pions is then
 \be
F_\nu(x,E_p)= \frac{2}{\lambda}\int_0^\lambda
F_\pi\left(\frac{E_\nu}{x},E_p\right)\frac{dx}{x},
 \ee
where $x=E_\nu/E_p$ and $\lambda= 0.427$.

\begin{figure}
\includegraphics[trim = 0mm 6mm 0mm 8mm, clip, width=9cm,angle=0]{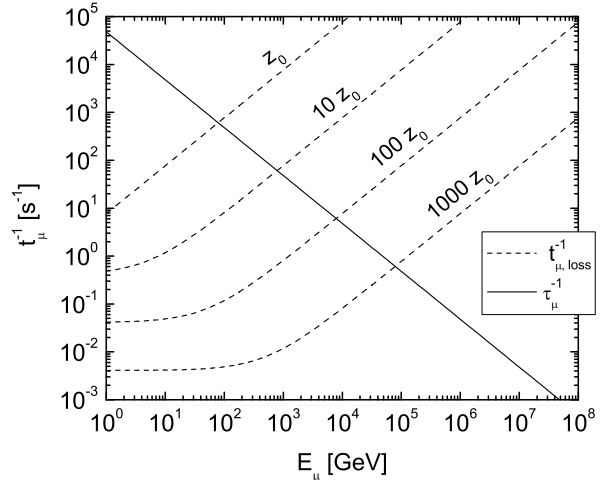}
\caption{Decay and loss rates for muons at different values of
$z_{\rm j}$.}\label{Figmuonloss}
\end{figure}

As it is the case for the pions, the muons also undergo synchrotron
and adiabatic losses. In this case, their mean lifetime is much
longer ($\tau_{\mu}= \tau_\mu^{(0)}\gamma_\mu $ with
$\tau_\mu^{(0)}=2.2\times 10^{-6}$ s). This implies that these
leptons will lose most of their energy before decaying, especially
at the inner parts of the jets, where most of the emission is
originated. This can be seen in Fig. \ref{Figmuonloss}, where we
show the decay rate $\tau_{\mu}^{-1}$ and the loss rate $t_{\mu,{\rm
loss}}^{-1}= t^{-1}_{\mu,{\rm sync}}+ t^{-1}_{\mu,{\rm adiab}}$ as a
function of the muon energy, for different values of $z_{\rm j}$.
According to this figure, muons with energies beyond $1 \ {\rm TeV}$
will be present in the jets only at $z_{\rm j}>10 z_0$. Hence, since
we are interested in neutrinos with energies $E_\nu> 1 \ {\rm TeV}$,
the neutrino emission is attenuated due to the synchrotron losses in
our model. The above equations show that to a good approximation we
can safely neglect the contribution from muon decays at high
energies.

\begin{figure}
\includegraphics[trim = 10mm 6mm 0mm 8mm, clip, width=9cm,angle=0]{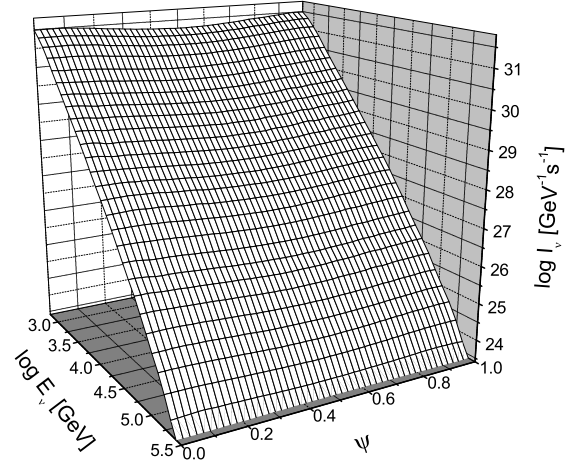}
\caption{Spectral intensity of neutrinos produced in the approaching
jet as a function of the precessional phase and the neutrino energy.}\label{FigInu-a}
\end{figure}

The neutrino emissivity can then be expressed as
  \be
\frac{dN_\nu(t,E_\nu,z_{\rm j})}{dE_\nu}= \int_{x_{\rm min}}^{x_{\rm
max}} \sigma_{pp}^{\rm inel}\left(\frac{E_\nu}{x}\right)
J_p\left(t,\frac{E_\nu}{x},z_{\rm j}\right)\times  \nonumber \\
F_{\nu}\left(x,\frac{E_\nu}{x}\right)  dx \\  \equiv
\left(\frac{z_0}{z_{\rm
j}}\right)^2\frac{d\tilde{N}_\nu(t,E_\nu)}{dE_\nu}.
  \ee
The total neutrino spectral intensity emitted in the jet thus reads
 \be
I_\nu(t,E_\nu)\simeq \frac{\dot{m}_{\rm j}z_0}{m_p v_{\rm
b}}\frac{d\tilde{N}_\nu(t,E_\nu)}{dE_\nu}.
 \ee
We show the  result obtained for neutrinos produced in the
approaching jet in Fig. \ref{FigInu-a}.

\section{Opacity to gamma-ray propagation}

The various mechanisms at work for gamma-ray absorption in the
microquasar SS433 have recently been studied in \citet{last}.
Absorption can occur via $\gamma\gamma$ interactions with ambient
soft photons and by $\gamma N$ interactions with disk and star
matter.

In the first case, absorption occurs through interactions with low
energy photons originated in the star and in the extended disk.
These take place in the pair creation channel,  and the
corresponding optical depth results from integration of
 \be
 d\tau_{\gamma\gamma}= (1-
\hat{e}_\gamma \cdot \hat{e}_{\rm ph}) n_{\rm ph}(E,\Omega')
\sigma_{\gamma\gamma} \ d\rho_\gamma \ dE  \ d \cos\theta' \ d\phi'
\label{dtau}
 \ee
as described in \citet{last}. Here, $d\rho_\gamma$ is the
differential path followed by the gamma ray, $E$ is the soft photon
energy, $\hat{e}_\gamma$ is the unit vector in the direction of the
gamma ray, and $\hat{e}_{\rm ph}= (\sin\theta' \cos\phi',\sin\theta'
\sin\phi',\cos\theta')$ is the vector directed along the direction
of the soft photons. The cross section for the process
$\gamma\gamma\rightarrow e^+e^-$ is
 \begin{multline}
\sigma_{\gamma\gamma}(E_\gamma,E)=\frac{\pi r_0^2}{2}(1-\xi^2)
\times \nonumber\\ \left[2\xi(\xi^2-2)+
(3-\xi^4)\ln\left(\frac{1+\xi}{1-\xi}\right)\right],
 \end{multline}
where $r_0$ is the classical electron radius and
 \be \xi=\left[1-\frac{2(m_e c^2)^2}{E_\gamma E(1-\hat{e}_\gamma \cdot
\hat{e}_{\rm ph})}\right]^{1/2}.
 \ee
The radiation density of soft photons, in units ${\rm cm}^{-3} {\rm
erg}^{-1}{\rm sr}^{-1}$, is $n_{\rm ph}(E,\Omega')= {2E^2}{(h
c)^{-3}( e^{E/kT}- 1)^{-1}}$ with $T=8500$ K for the starlight
photons and with $T=21000$ K for the UV photons from the extended
disk. The mid-IR emission is characterized by a radiation density
$n_{\rm ph}(E,\Omega')\approx {F_{\rm IR}d^2}/({hc E \pi
r_\gamma^2\cos(0.62\pi)})$, with $F_{\rm IR}= 2.3\times
10^{-23}(\lambda/\mu{\rm m})^{-0.6}$ for $ 2 \ \mu{\rm m}<\lambda<12
\ \mu{\rm m}$ \citep{Fuchs05}.

As for the absorption due to interactions with matter, the important
effects are photopion production $\gamma N \rightarrow \pi^i \gamma$
and photopair production $\gamma N \rightarrow N e^+e^-$, where $N$
represents a nucleon. This last effect has not been previously
considered, so it will be taken into account in the present work. We
assume that the star has a matter density
$$\rho_\star(r)=
\frac{M_\star}{4\pi R_\star r^2} \Theta(r-R_\star),$$ where $r$ is
the distance from the gamma-ray position to the center of the star.
For the extended disk, we consider that matter density is given by
$$\rho_{\rm w}(r_\gamma)= \frac{\dot{M}_{\rm w}}{v_{\rm w}
\Delta\Omega r_\gamma^2},$$ for $60^\circ<\theta_Z<120^\circ$, where
$\theta_Z$ is the polar angle in a coordinate system with its
$Z$-axis directed along the approaching jet axis (for details, see
Reynoso et al. 2008). The $\gamma N$ contribution to the optical
depth is
 \be
\tau_{\gamma N}(\vec{z}_{\rm j})= \int_0^\infty \sigma_{\gamma N}
\frac{\left( \rho_\star+ \rho_{\rm w }\right)}{m_p} d\rho_\gamma,
\label{tauGN}
 \ee
where $\sigma_{\gamma N}= \sigma_{p\gamma}^{(\pi)}+
\sigma_{p\gamma}^{(e)}$ can be obtained from equations
(\ref{sigphotopion}) and (\ref{sigphotopair}).

\begin{figure}
\includegraphics[trim = 5mm 6mm 0mm 8mm, clip, width=9cm,angle=0]{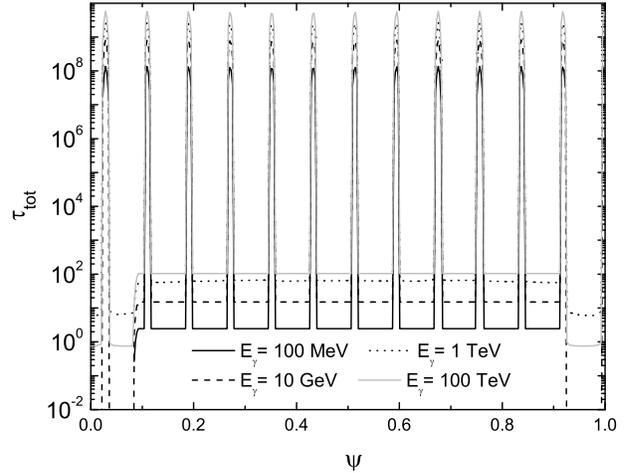}
\caption{Total optical depth as a function of the precessional phase
for gamma rays of different energies originated at the injection
point $z_0$ of the approaching jet.}\label{Figtautotz0-a}
\end{figure}

\begin{figure}
\includegraphics[trim = 20mm 6mm 0mm 8mm, clip, width=9cm,angle=0]{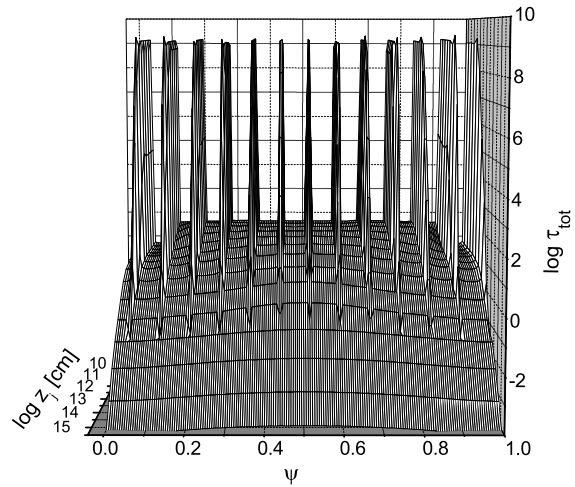}
\caption{Total optical depth as a function of the precessional phase
and $z_{\rm j}$ for gamma-rays with $E_\gamma=1$ TeV originated in
the approaching jet.}\label{Figtau1TeVzj}
\end{figure}

In Fig. \ref{Figtautotz0-a} we show the total optical depth as a
function of the precessional phase for different energies of gamma
rays originated at the injection point $z_0$ of the approaching jet.
A very similar result is obtained the base of the receding jet,
since $z_0$ is much smaller than any other size scale of the system.
We clearly see the peaks of extreme absorption produced every time
the star eclipses the emission region. The dependance of the total
optical depth on the distance to the black hole $z_{\rm j}$ is also
shown in Fig. \ref{Figtau1TeVzj} for gamma rays of energy
$E_\gamma=1$ TeV coming from the approaching jet.

\section{Neutrino and gamma-ray detectability}\label{neutrinos}
The spectral intensities shown in Figs. \ref{FigIga-a} and
\ref{FigInu-a} were obtained for $q_{\rm rel}= 10^{-4}$ without
considering absorption. However, it is of course necessary to
include the absorption effects to see how they affect the produced
fluxes that may arrive to the Earth.

The differential gamma-ray flux to be observed from each jet can be
obtained as
 \be
\frac{d\Phi_\gamma(t,E_\gamma)}{dE_\gamma}=\frac{1}{4\pi
d^2}\int_{z_0}^{z_1}\pi(z_{\rm j}\tan \xi)^2 \ n_p\times
\\ \frac{dN_\gamma(t,E_\gamma,z_{\rm j})}{dE_\gamma} e^{-\tau_{\rm
tot}(t,E_\gamma,z_{\rm j})}dz_{\rm j}.
 \ee

\begin{figure}
\includegraphics[trim = 10mm 6mm 0mm 8mm, clip, width=9cm,angle=0]{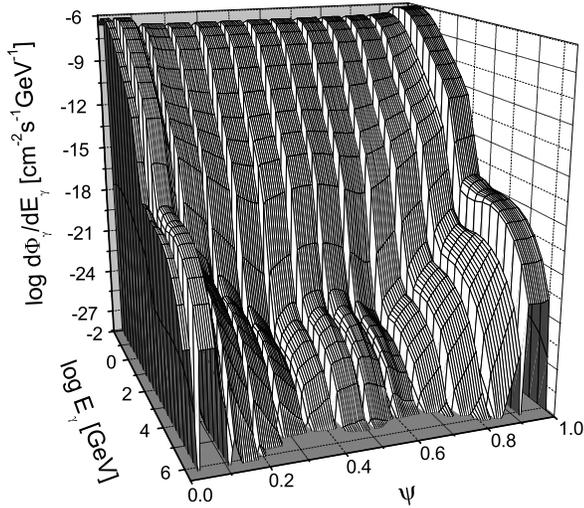}
\caption{Differential gamma-ray flux arriving at Earth as a function
of precessional phase and energy.}\label{FigFgaboth}
\end{figure}

We show in Fig. \ref{FigFgaboth} the joint contribution of both jets
to the differential gamma-ray flux considering $q_{\rm rel}=
10^{-4}$. Here, the absorption effects  and the precessional phase
behavior have  become manifest in the spectrum (c.f. Fig.
\ref{FigIga-a}).

As for neutrinos, although they undergo only weak interactions, we
can estimate the corresponding neutrino optical depth using an
expression analogous to (\ref{tauGN}),
 \be
 \tau_{\nu N}(\vec{z}_{\rm j})= \int_0^\infty
 \sigma_{\nu N} \frac{\left( \rho_\star+ \rho_{\rm w }\right)}{m_p}d\rho_\gamma.
 \ee
Here the total $\nu_\mu N$ cross section can be approximated for
$E_\nu>1$ TeV as
 $
\sigma_{\nu N}( E_\nu) \approx 10^{\alpha_\nu(E_\nu)}
 $${\rm cm}^2$, with
 \begin{multline}
 \alpha_\nu(E_\nu)=-38.42+
1.46\log{\left(\frac{E_\nu}{\rm GeV}\right)}-
0.116\log^2{\left(\frac{E_\nu}{\rm GeV}\right)}+
\\ 0.0041\log^3{\left(\frac{E_\nu}{\rm GeV}\right)}.\nonumber
 \end{multline}

\begin{figure}
\includegraphics[trim = 10mm 6mm 0mm 8mm, clip, width=9cm,angle=0]{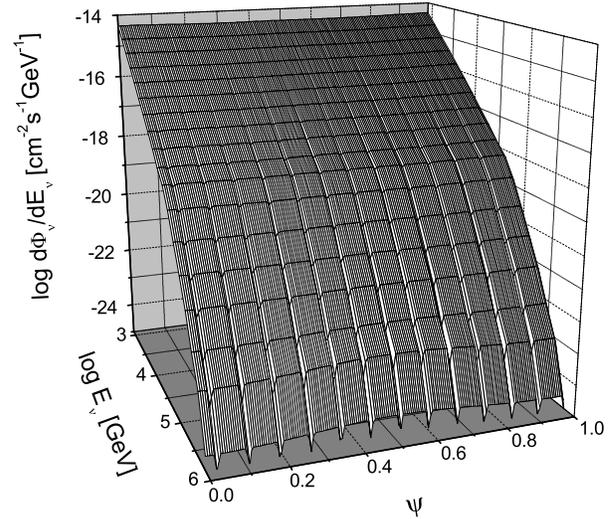}
\caption{Differential neutrino flux arriving at Earth as a function
of precessional phase and energy.}\label{FigFnuboth}
\end{figure}

The differential neutrino flux arriving to Earth can therefore be
estimated as
 \begin{multline}
\frac{d\Phi_\nu(t,E_\nu)}{dE_\nu}=\frac{1}{8\pi
d^2}\int_{z_0}^{z_1}\pi(z_{\rm j}\tan \xi)^2 \ n_p\times
\\ \frac{dN_\nu(t,E_\nu,z_{\rm j})}{dE_\nu} e^{-\tau_{\nu N}(t,E_\nu,z_{\rm j})}dz_{\rm
j}.
 \end{multline}
Notice that an additional $1/2$ factor has been put in order to take
into account the reduction in the muon neutrino flux due to flavor
oscillations over astrophysical distances (e.g. Athar et al. 2005).
 The result for the differential neutrino flux
from the two jets with $q_{\rm rel}=10^{-4}$ is shown in Fig.
{\ref{FigFnuboth}} as a function of precessional phase and neutrino
energy.

\begin{figure}
\includegraphics[trim = 0mm 6mm 0mm 8mm, clip, width=9cm,angle=0]{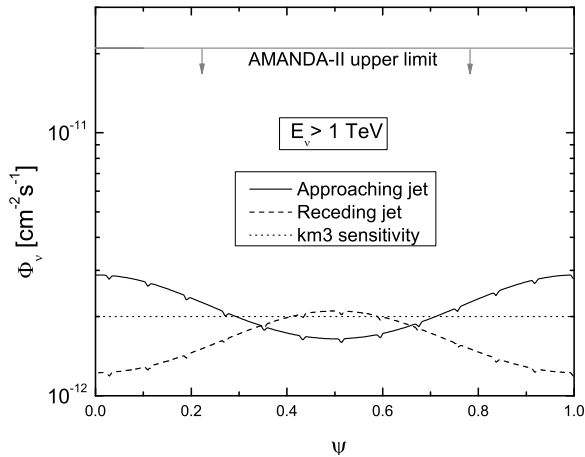}
\caption{Neutrino fluxes arriving at Earth with $E_\nu>1$ TeV as a
function of the precessional phase. The contributions of the two
jets are shown separately, solid line for the approaching jet and
dashed line for the receding one. The upper limit from AMANDA-II
data and the expected km$^3$ sensitivity are shown in grey solid
line and dotted line respectively.}\label{Fignut}
\end{figure}

We can now integrate the differential fluxes in energy to appreciate
the precessional dependance of the signals, $$\Phi(\psi)=
\int_{E^{\rm(min)}}^{E^{\rm(max)}}\frac{d\Phi}{dE}dE.$$ It is
interesting to explore the energy ranges that are expected to be
covered with different instruments. In the case of neutrinos with
$E_\nu>1$ TeV, in view of possible detection with IceCube, we obtain
the flux shown in Fig. \ref{Fignut}, where the contributions from
the two jets appear separately. It is also included in that figure
the value of the upper limit that can be extracted from AMANDA-II
data, $\Phi_\nu^{\rm(lim)}(E_\nu>1{\ \rm TeV})= 2.1\times
10^{-11}{\rm cm}^{-2}{\rm s}^{-1}$ \citep{Halzen06}, which agrees
with recently published experimental results \citep{AMANDA07}. The
expected sensitivity for a km$^3$ neutrino telescope such as
IceCube, $\Phi_\nu^{\rm(km3)}(E_\nu>1{\ \rm TeV})\approx 2\times
10^{-12}{\rm cm}^{-2}{\rm s}^{-1}$ for three years of operation, is
also shown in Fig. \ref{Fignut} \citep{Halzen06,nemo,nemo2}.

\begin{figure}
\includegraphics[trim = 0mm 6mm 0mm 8mm, clip, width=9cm,angle=0]{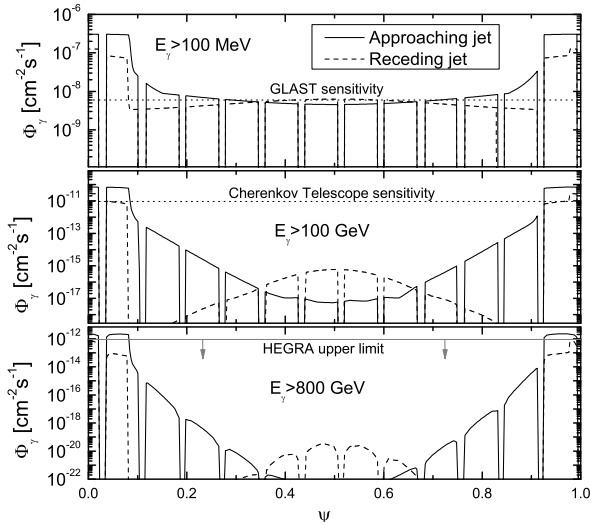}
\caption{Gamma-ray fluxes arriving at Earth as a function of the
precessional phase for $100 \ {\rm MeV}<E_\gamma<300$ GeV in the
\textit{upper panel}, for $E_\gamma>100$ in the {\it middle panel},
and for $E_\gamma>800$ GeV in the \textit{lower panel}. The
contributions of the two jets are shown separately: \textit{solid
line} for the approaching jet and \textit{dashed line} for the
receding one.}\label{Figgat}
\end{figure}

As for gamma rays, the sensitivity expected for GLAST is
$\Phi_\gamma^{\rm (GLAST)}\simeq 6\times10^{-9}{\rm cm}^{-2}{\rm
s}^{-1}$ at $100 \ {\rm MeV}<E_\gamma<300$ GeV\footnote{See the
official NASA GLAST webpage,
{http://glast.gsfc.nasa.gov/science/instruments/table1-1.html}},
while for Cherenkov telescopes such as VERITAS and MAGIC II we
consider a sensitivity $\Phi_\gamma^{\rm (Cher)}\simeq
9\times10^{-12}{\rm cm}^{-2}{\rm s}^{-1}$ for energies
$E_\gamma>100$ GeV \citep{Cherenkov}. We show in Fig. \ref{Figgat}
the integrated gamma-ray fluxes expected for $q_{\rm rel}=10^{-4}$
in the mentioned energy ranges, as compared with the respective
sensitivities. There, the contributions from the two jets are
plotted separately. We also include in the lower panel the results
corresponding to $E_\gamma>800$ GeV, since they allow a comparison
with the upper limit given by HEGRA for that energy range
\citep{HEGRA}, $\Phi_\gamma^{\rm(lim)}(E_\gamma>0.8 \ {\rm TeV})=
8.93 \times 10^{-13}{\rm cm}^{-2}{\rm s}^{-1}$.

According to our model, the most favorable range of precessional
phases for the gamma-ray detection is found to be between
$\psi\lesssim0.09$ and $\psi\gtrsim0.91$, as it can be seen from
Fig. \ref{Figgat}. This corresponds to the cases when the extended
disk is so much open in our direction that the gamma rays emitted at
the innermost regions can escape without having to travel through
the thick extended disk, undergoing only $\gamma\gamma$ absorption.

However, since the observations by HEGRA were not performed
exclusively at the dates of favorable precessional phases
\citep{HEGRA}, we compare the value of the HEGRA cut with that of
the averaged flux over one precession cycle, according to our model.
Thus, the implied maximum value allowed by the HEGRA limit for the
free parameter $q_{\rm rel}$ is
$$q_{\rm rel}^{\rm(max)}=
10^{-4}\frac{\Phi_\gamma^{\rm(lim)}}{\langle\Phi_\gamma\rangle}\approx
2.9\times 10^{-4},$$which is below the maximum value allowed by the
AMANDA II limit.




Given that this parameter can be linearly factored out in our
equations provided that $q_{\rm rel}\ll 1$, which is guaranteed for
the allowed cases, we can plot the predictions for the different
averaged fluxes as a function of this parameter and compare them
with the expected sensitivities for the different instruments.



\begin{figure}
\includegraphics[trim = 0mm 6mm 0mm 0mm, clip, width=9cm,angle=0]{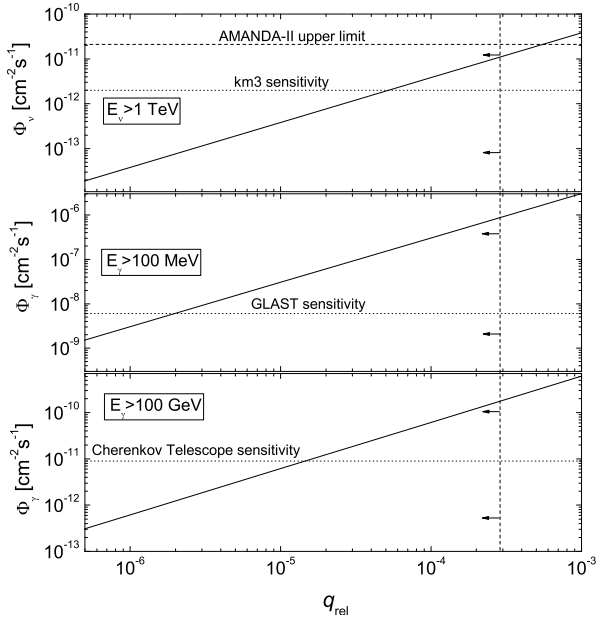}
\caption{Dependance on $q_{\rm rel}$ of the different flux
predictions: average neutrino flux (\textit{upper panel}) and
average gamma-ray flux for $100 \ {\rm MeV}<E_\gamma<300$ GeV ({\it
middle panel}) and for $E_\gamma>100$ GeV ({\it lower panel}).
The vertical line indicates the maximum possible value of $q_{\rm
rel}$ derived from HEGRA observations. The corresponding
sensitivities are also show (see the text for
details).}\label{Figvaryqrel}
\end{figure}

This is done in Fig. \ref{Figvaryqrel}, where the mean neutrino flux
as a function of $q_{\rm rel}$ is shown in the upper panel as
compared with the AMANDA-II upper limit and with the expected
sensitivity for km$^3$ neutrino telescopes such as IceCube. The
gamma-ray fluxes for $100 \ {\rm MeV}<E_\gamma<300$ GeV and for
$E_\gamma>100$ GeV averaged in the above mentioned range of
favorable precessional phases are shown in the middle and lower
panels of the same figure, as a function of $q_{\rm rel}$. The
vertical line indicates the bound $q_{\rm rel}^{\rm(max)}$ derived
from HEGRA observations, which is clearly more restrictive than the
one implied by the AMANDA-II upper limit.

\section{DISCUSSION and SUMMARY}

We have studied the high-energy emission originated in the dark jets
of the microquasar SS433. A small fraction of its particle contents
are relativistic protons that collide with the cold ions within the
jets, producing gamma rays and neutrinos after pion decay processes.
%
%
We found that up to distances $\sim 10^{12}{\rm cm}$ from the black
hole, protons with energies below $\sim 3\times 10^{6}$GeV will cool
dominantly via $pp$ interactions. The ratio of the power carried by
 relativistic protons to the total kinetic power of the jet,
$q_{\rm rel}$, was kept as a free parameter of the model (for
illustrative purposes we have used $q_{\rm rel}= 10^{-4}$ in the
figures).

We have calculated the spectrum of gamma rays and high-energy
neutrinos based on the formulae given by \citet{Kelner06}. We have
considered the cooling of the charged pions and muons produced, and
we have found that the high-energy neutrino emission is attenuated
by synchrotron losses. Adding the contribution from both jets, we
have obtained the total gamma-ray spectral intensity of SS433. The
gamma radiation will be largely absorbed while leaving the inner
regions the system by means of several processes. This will mainly
occur through interaction with matter of the star and extended disk,
leading to significant photopair production.  UV and mid-IR photons
originated in the extended disk are also expected to cause important
absorption via $\gamma\gamma$ annihilations.

The total optical depth is found to depend on the precessional phase
in such a way that when the approaching jet is pointing away from
the Earth, at $\psi\sim 0.5$, the extended disk blocks the emitting
region and the absorption is strongest. In particular, in the range
of precessional phases between $\psi \gtrsim 0.91$ and $\psi
\lesssim 0.09$, the gamma rays originated at the base of the jets
will travel to the Earth without having to pass through the
equatorial disk. With the conservative assumption that this
outflowing disk presents a large half opening angle $\alpha_{\rm
w}=30^\circ$, the mentioned range of favorable precessional phases
gives a total of $\sim 29$ d for optimal detectability. Since,
according to \citet{Gies02b}, $\psi=0$ occurred on 2002 June 5, it
follows that the next upcoming opportunities to achieve detection
will be centered around the following dates every $162$ d: 2008
August 20, 2009 January 29, 2009 July 10, etc. As mentioned above,
the exact duration along which the favorable conditions may hold,
depends on the half opening angle of the extended disk, which might
be smaller than what was assumed. In that case, the observational
window would be broader

The observations by HEGRA imply a maximum value for the free
parameter $q_{\rm rel}$ in $\sim 3 \times 10^{-4}$. Given the
expected sensitivity of the next km$^3$ neutrino telescopes
generation, it will be possible to test our model 
down to values $\sim 5\times 10^{-5}$ in three years of operation.
An extended range of this parameter will be probed by gamma-ray
observations with GLAST and Cherenkov telescopes, especially if
performed on the favorable dates.

We conclude that its dark jets can be the possible site for both
gamma-ray and neutrino production in SS433. Since most of the
high-energy flux is generated in the inner jets, gamma-ray
absorption will make detection with Cherenkov telescopes like
VERITAS and MAGIC-II difficult but not impossible if attempted at
the favorable dates when the approaching jet is closest to the line
of sight. Actually, there are much better prospects for gamma-ray
observation with GLAST and neutrino detection with IceCube also
seems promising. 
The determination of the gamma-ray to neutrino flux ratio would
allow to estimate $q_{\rm rel}$ unambiguously, yielding crucial
information about acceleration mechanisms taking place in the jets.

\section*{Acknowledgments}
We thank an anonymous referee for his/her constructive comments on
this work. G.E.R. is supported by the Argentine agencies CONICET
(PIP 452 5375) and ANPCyT (PICT 03-13291 BID 1728/OC-AR). additional
support is provided by the Ministerio de Educaci\'on y Ciencia
(Spain) under grant AYA2007-68034-C03-01, FEDER funds. H.R.C. is
supported by FUNCAP, Brazil, and M.M.R. is supported by CONICET,
Argentina. M.M.R. is also thankful to O.A. Sampayo for very useful
discussions.



\end{document}